\documentclass{article}

\usepackage[preprint]{neurips_2026}
\usepackage{graphicx}
\usepackage{amsmath,amssymb}
\usepackage{hyperref}    
\usepackage{tipa}        
\usepackage{multirow}
\usepackage{array}
\usepackage{subcaption}
\usepackage{caption}
\usepackage{kotex}
\usepackage{wrapfig}

\usepackage[utf8]{inputenc} % allow utf-8 input
\usepackage[T1]{fontenc}    % use 8-bit T1 fonts
\usepackage{hyperref}       % hyperlinks
\usepackage{url}            % simple URL typesetting
\usepackage{booktabs}       % professional-quality tables
\usepackage{amsfonts}       % blackboard math symbols
\usepackage{nicefrac}       % compact symbols for 1/2, etc.
\usepackage{microtype}      % microtypography
\usepackage{xcolor}         % colors

\title{Mask2Flow-TSE: Two-Stage Target Speaker Extraction with Masking and Flow Matching}

\author{%
  Junwon Moon\textsuperscript{1}, Seungbeom Kim\textsuperscript{1}, Hansol Park\textsuperscript{1}, Hyunjin Choi\textsuperscript{1},  Hoseong Ahn\textsuperscript{1}, \\ \textbf{Heeseung Kim}\textsuperscript{2}, \textbf{Kyuhong Shim}\textsuperscript{1} \\
  \textsuperscript{1}Sungkyunkwan University
  \textsuperscript{2}University of Seoul\\
  \texttt{\{mppn98, khshim\}@skku.edu} \\
}

\begin{document}

\maketitle
\begin{abstract}
Target speaker extraction (TSE) extracts the target speaker's voice from overlapping speech given a reference utterance.
Existing masking-based approaches are lightweight and effective but suffer from an inability to synthesize missing content, leading to degraded perceptual quality.
On the other hand, recent generative TSE models typically synthesize high-quality speech with diffusion,
but require numerous iterative steps resulting in high computational costs and latency.
We propose \textbf{Mask2Flow-TSE}, a two-stage framework combining the strengths of both paradigms.
We introduce the deletion/insertion (D/I) proportion, an analytical tool that reveals early flow steps predominantly remove signal components rather than synthesize them.
Based on this finding, we decouple deletion from insertion: a masking-based module handles the deletion-dominant early steps, while a single flow-matching step performs the remaining insertion for high-quality reconstruction.
Specifically, the first stage uses lightweight convolution for the masking module, while the second stage employs a Diffusion Transformer (DiT) adapted for TSE with speaker conditioning.
Unlike prior approaches that start from Gaussian noise, our method starts from the masked spectrogram, enabling high-quality reconstruction in a single inference step.
Experiments show that Mask2Flow-TSE produces high-quality extractions with only 85M parameters and one-step inference, while preserving clean single-speaker inputs with minimal degradation.
\end{abstract}

\section{Introduction}
\label{sec:intro}

In real-world scenarios, automatic speech recognition (ASR) systems often
struggle when multiple speakers talk simultaneously. This challenge, commonly
referred to as the cocktail-party problem~\cite{cherry1953some}, has long been a
fundamental obstacle in speech processing~\cite{watanabe2020chime,yu2022m2met}. Target speaker extraction (TSE)
addresses this by extracting only the desired speaker's voice from a speech mixture (i.e., overlapped speech)  while
suppressing interfering speakers and background noise~\cite{ge2020spex+, vzmolikova2019speakerbeam}.
TSE plays a critical role
in improving the robustness of downstream applications such as ASR, hearing aids,
and telecommunication systems, where accurate extraction under strict latency
and model size constraints is essential~\cite{borsdorf2024wtimit2mix, polok2025target}.

Existing TSE approaches can be broadly categorized into discriminative and
generative methods. Among discriminative methods, masking-based models~\cite{nguyen2024convoifilter, wang2020voicefilter, wang2018voicefilter}
apply a soft mask to the input mixture to retain only the target speaker's voice.
These models are typically lightweight and fast. However, they can only remove components from the mixture; therefore, overly suppressed target cannot be recovered.
In contrast, generative methods such as diffusion models~\cite{kamo2023target, scheibler2023diffusion}
directly synthesize the target speech, which can better restore severely degraded
regions.
However, in addition to their large model sizes, these models typically
require many iterative refinement steps, leading to slow inference~\cite{wang2025solospeech,
wang2025metis}.
Recently, flow matching~\cite{lipman2022flow, liu2022flow, navon2025flowtse} has reduced inference steps compared to diffusion, but it still requires multi-step sampling for high-quality output~\cite{wang2025flowse}.
As a result, it remains difficult for a single method to
achieve low-latency inference, compact model size, and high extraction quality
at the same time, even though they are essential for TSE to be a practical ASR
front-end.

\begin{figure}[t!]
  \centering
  \includegraphics[width=1.0\linewidth]{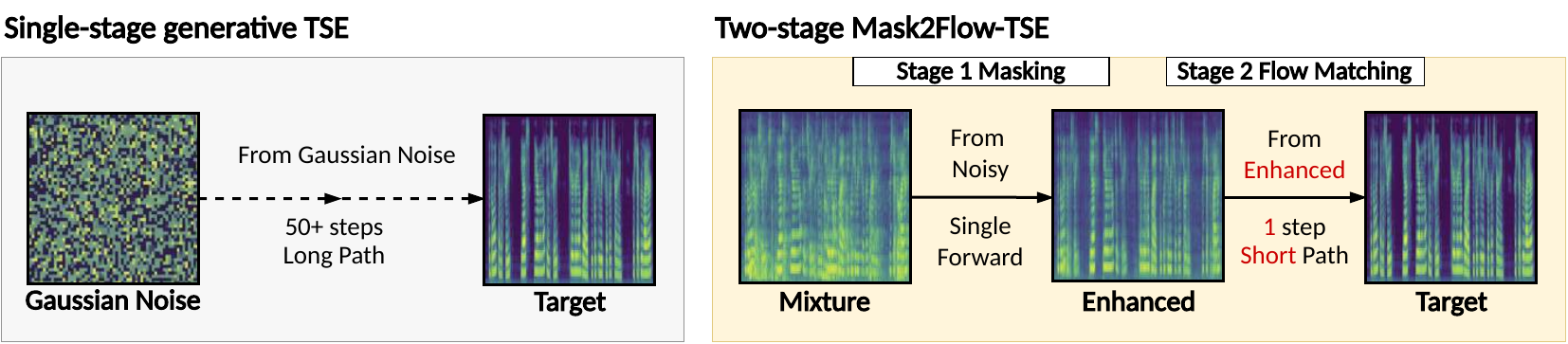}
  \caption{Single-stage generative TSE vs. our two-stage Mask2Flow-TSE. Single-stage
  approaches start from Gaussian noise and require many iterative steps, while
  our method starts from the masking-enhanced spectrogram, reducing inference to
  only 1 step.}
  \label{fig:two_stage_paradigm}
\end{figure}

In this paper, we propose \textbf{Mask2Flow-TSE}, a two-stage framework that bridges
masking-based methods and flow-matching models for TSE. In the first stage, a lightweight
masking module performs coarse extraction, and in the second stage, flow matching refines
the output toward the target speech. Figure~\ref{fig:two_stage_paradigm}
illustrates the key idea. Combining masking with flow matching for
TSE has not been explored, yet the two paradigms are naturally complementary:
masking efficiently handles deletion of interfering components, and flow matching
provides the generative ability to insert spectral details that masking cannot recover.
Beyond this, the proposed two-stage design offers an additional efficiency benefit: explicit masking serves as an effective initialization for flow matching, making single-step Euler inference possible from the masked spectrogram rather than from Gaussian noise.

Our design is motivated by the empirical observation that flow-based TSE models
exhibit masking-like behavior during the early inference steps. To validate
this, we introduce the delete--insert (D/I) proportion, which quantifies how each inference
step cumulatively modifies the spectrogram relative to the input mixture.
Our analysis reveals that deletion dominates the early steps, closely resembling discriminative masking.
This motivates replacing the early flow steps with a lightweight masking module, substantially reducing the computational burden.
Meanwhile, we observe that reconstructing the target speech requires substantial insertion that subtractive masking alone cannot provide; this motivates the second generative refinement stage.
Finally, we show that even a single flow step is sufficient when combined with masking as the first stage.

\vspace{0.1cm}
\noindent
The main contributions of this work are as follows:
\begin{itemize}
  \item We propose Mask2Flow-TSE, the first target speaker extraction framework that combines discriminative
    masking with generative flow matching.

  \item We introduce a D/I proportion analysis that reveals the deletion-dominant nature of early flow-based inference in TSE. This finding motivates our two-stage design in which a lightweight masking module replaces the early flow steps to reduce computational complexity.

  \item Mask2Flow-TSE achieves the lowest WER under noisy conditions among the evaluated models while preserving competitive speech quality, with only 85M parameters and a single inference step. Notably, achieving comparable speech recognition
  performance without Mask2Flow-TSE requires an approximately 10$\times$ larger ASR model.
\end{itemize}
\section{Related Work}\label{sec:related}

\subsection{Masking-based TSE}\label{ssec:related_voicefilter}
Masking-based models first predict a soft mask from the mixture conditioned on a speaker embedding, and then apply it to suppress non-target speech while retaining the target speaker's voice.
By multiplying the mask with the input mixture spectrogram, regions dominated by interference are attenuated toward zero.
Representative masking-based models include VoiceFilter~\cite{wang2018voicefilter} and its lightweight variant VoiceFilter-Lite~\cite{wang2020voicefilter},
employing CNN and LSTM architectures for mask estimation.
More recently, ConVoiFilter~\cite{nguyen2024convoifilter} replaces LSTM and CNN with Conformer~\cite{gulati2020conformer} blocks to effectively capture contextual information.
Additionally, attention-enhanced temporal convolutional networks have been introduced~\cite{wang2024target}, and SEF-Net~\cite{zeng2023sef} employs cross-attention to fuse reference and mixture encodings without explicit speaker embeddings.

This discriminative approach enables lightweight and efficient architectures, making such models attractive for real-time applications.
However, because a multiplicative mask is constrained to $[0, 1]$, masking is inherently a deletion-only operation; it can suppress interfering components but cannot increase energy beyond the input mixture spectrogram.
Consequently, when the target speech is over-suppressed or strongly masked by interference, masking alone may lead to degraded speech quality and recognition performance.

\subsection{Generative Models}\label{ssec:related_generative}

Generative approaches have emerged as alternatives to discriminative methods, directly synthesizing target speech from a learned distribution rather than filtering the input mixture.
Recently, diffusion-based models~\cite{ho2020denoising} such as DiffSep~\cite{scheibler2023diffusion} and DiffTSE~\cite{kamo2023target} apply score-based methods to speech separation and TSE, respectively.
DDTSE~\cite{zhang2024ddtse} and SoloSpeech~\cite{wang2025solospeech} extend diffusion-based TSE frameworks, achieving improved perceptual quality and intelligibility.
DiffSpEx~\cite{nguyen2023conditional} and AVDiffuSS~\cite{lee2024seeing} also extend conditional diffusion to single-channel and audio-visual TSE settings, respectively.
Despite their superior perceptual quality, diffusion-based methods generally require iterative sampling starting from Gaussian noise, resulting in slow inference.

Flow matching~\cite{lipman2022flow, liu2022flow} offers a more efficient generative framework by learning straight transport paths between distributions, reducing the number of integration steps compared to diffusion models.
For example, FlowTSE~\cite{navon2025flowtse} applies conditional flow matching to TSE in the mel-spectrogram domain.
However, most existing flow-based TSE methods still generate target speech entirely from Gaussian noise, requiring the model to handle the full transformation from noise to clean speech.

In speech enhancement (SE), several works have aimed to reduce the inference cost of iterative generation.
FlowSE~\cite{wang2025flowse} applies flow matching to SE in the mel-spectrogram domain, substantially reducing latency compared to diffusion-based models.
FlowAVSE~\cite{jung2024flowavse} extends this to the audio-visual setting. 
Despite this progress in SE, similar improvements have yet to be demonstrated for TSE.

\subsection{Two-stage Models}\label{ssec:related_storm}
Two-stage approaches that combine discriminative and generative models have recently
shown strong performance in SE~\cite{dhyani2025high}.
For example, Gesper~\cite{liu2023gesper} regenerates damaged speech and then applies discriminative SE, reducing the risk of excessive suppression.
Hi-ResLDM~\cite{dhyani2025high} separates each stage's goal; the former discriminative stage increases the SNR to remove noise, while the latter diffusion stage focuses on restoring speech.
StoRM~\cite{lemercier2023storm} attaches a discriminative model to the front of a diffusion model, starting generation from noise-perturbed discriminative output.
However, these methods \textit{do not address} target speaker extraction (TSE), where the system must extract a reference-specified speaker from competing speakers rather than generally enhance corrupted speech. 
In addition, their generative stages still require multiple sampling steps, even when paired with a discriminative front-end.

To the best of our knowledge, Mask2Flow-TSE is the first framework to combine discriminative masking with generative flow matching for TSE.
By assigning distinct roles to masking and flow matching, Mask2Flow-TSE combines the
efficiency of discriminative models with the reconstruction capability of generative
models, requiring only a single Euler step for refinement.

\section{Preliminaries}\label{sec:prelim}

\subsection{Target Speaker Extraction}\label{ssec:problem}

Given a mixture signal $x = y + z$, where $y$ is the target speaker's
clean speech and $z$ contains one or more interfering speakers and
possibly background noise
, target speaker extraction (TSE) aims to
isolate $y$ from $x$.
To identify which speaker to extract, TSE systems rely on a short reference utterance from the target speaker.
A speaker encoder processes this reference to produce a fixed-dimensional embedding $d$, commonly referred to as a d-vector~\cite{wan2018generalized}.
This embedding captures the speaker's voice characteristics and guides the extraction network to focus on the corresponding speaker in the mixture.
Formally, given an input mixture $x$ and speaker embedding $d$, the TSE model $F_\theta$ estimates the clean target speech:
\begin{equation}
    \hat{y} = F_\theta(x, d),
    \label{eq:tse}
\end{equation}
where $\hat{y}$ denotes the predicted clean speech of the target speaker.

\subsection{Flow Matching}\label{ssec:flow_matching}

Flow matching learns a velocity field that directly transports samples from a source distribution to a target distribution via an ordinary differential equation (ODE).
Given a source distribution $p_0$ and a target distribution $p_1$, flow matching defines a probability path $p_t$ for $t \in [0, 1]$ that interpolates between them, with $x_t \sim p_t$ denoting a sample along this path.
The transformation is governed by the ODE:
\begin{equation}
    \frac{dx_t}{dt} = v_\theta(x_t, t),
    \label{eq:ode}
\end{equation}
where $v_\theta$ is a neural network that predicts the velocity field.
A general formulation of the interpolation path is:
\begin{equation}
    x_t = (1 - t)x_0 + tx_1 + \sigma_t \epsilon,
    \label{eq:interpolation_general}
\end{equation}
where $x_0 \sim p_0$, $x_1 \sim p_1$, $\epsilon \sim \mathcal{N}(0, I)$, and $\sigma_t$ controls the noise level.
Under the optimal transport formulation with $\sigma_t = 0$, this simplifies to a straight line:
\begin{equation}
    x_t = (1 - t)x_0 + tx_1,
    \label{eq:interpolation}
\end{equation}
yielding a constant velocity $x_1 - x_0$ along the trajectory.
For this straight-line path, the training objective becomes:
\begin{equation}
\mathcal{L}_{\text{flow}}(\theta) = \mathbb{E}_{t,x_0,x_1} ||v_\theta(x_t,t)-(x_1-x_0)||^2
\label{eq:loss}
\end{equation}
At inference, the learned velocity field is integrated from $t=0$ to $t=1$ using numerical ODE solvers.
This formulation, known as rectified flow matching~\cite{liu2022flow}, enables accurate generation with significantly fewer integration steps.

\section{Mask2Flow-TSE}\label{sec:method}

\subsection{Motivation: Why Two Stages?}\label{ssec:motivation}

\begin{figure*}[t!]
  \centering
  \includegraphics[width=1.0\linewidth]{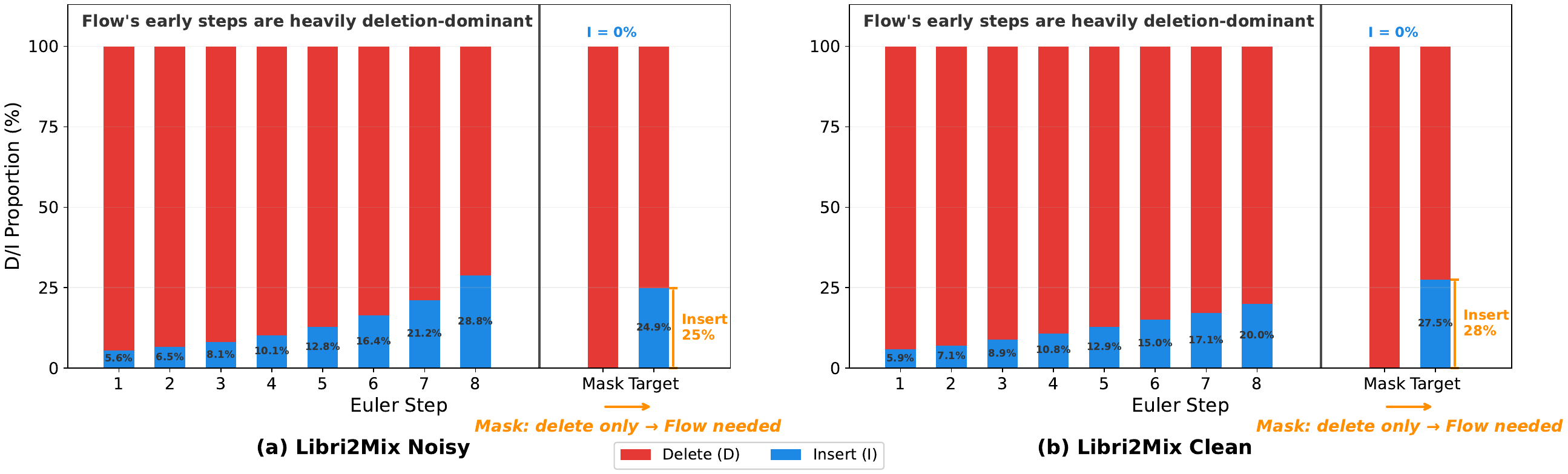}
  \caption{Delete--insert (D/I) proportion of a flow-only TSE model across 8 Euler steps.
  Each bar decomposes the total energy change from the input mixture
  into \textit{Delete} (D, red) and \textit{Insert} (I, blue).
  \textit{Mask}: a separately trained discriminative masking model.
  \textit{Target}: ground-truth clean speech.
  }
  \label{fig:motivation}
\end{figure*}

We hypothesize that flow-based TSE models inherently perform masking-like deletion during their early inference steps before generating fine spectral details.
To verify this, we analyze the cumulative behavior of an 8-step flow TSE model using a new analysis tool: \textit{delete--insert (D/I) proportion}.

For step~$k$, we compare the intermediate output $\hat{X}_k$ against the input mixture $X$ in the mel-spectrogram domain.
At each time--frequency bin $(t,f)$, we compute the change of the energy:
\begin{equation}
  \Delta_{t,f}^{(k)} = \hat{X}_k(t,f) - X(t,f).
  \label{eq:delta}
\end{equation}
The total deletion and insertion energies are then defined as:
\begin{equation}
  D^{(k)} = \sum_{\Delta_{t,f}^{(k)} < 0} |\Delta_{t,f}^{(k)}|, \quad
  I^{(k)} = \sum_{\Delta_{t,f}^{(k)} > 0} \Delta_{t,f}^{(k)},
  \label{eq:di_energy}
\end{equation}
and the D/I proportion is $D^{(k)} / (D^{(k)} + I^{(k)}) \times 100$.
Note that for this computation, the mel-spectrograms are scaled to $[0, 1]$ and the flow trajectory is initialized with the mixture prior.
Figure~\ref{fig:motivation} shows the average D/I proportion on the Libri2Mix test set, alongside a separately trained discriminative masking model (\textit{Mask}) and the ground-truth clean speech (\textit{Target}).

\noindent We make a key observation: every flow step is heavily deletion-dominant, particularly in the early steps.
As shown in Figure~\ref{fig:motivation}(a), step~1 exhibits D\,$\approx$\,94\% with only I\,$\approx$\,6\%; even step~8 remains at D\,$\approx$\,71\%.
This implies that the model spends a large portion of its inference budget suppressing interference, which is an operation that discriminative masking can accomplish in a single forward pass.
This motivates explicitly separating the deletion and insertion roles into two dedicated stages.

\subsection{Two-Stage Framework}\label{ssec:overall}

Mask2Flow-TSE is a two-stage framework that replaces the deletion-dominant operations with an explicit masking stage and performs insertion-dominant refinement via flow matching.

\begin{figure*}[t!]
  \centering
  \includegraphics[width=1.0\linewidth]{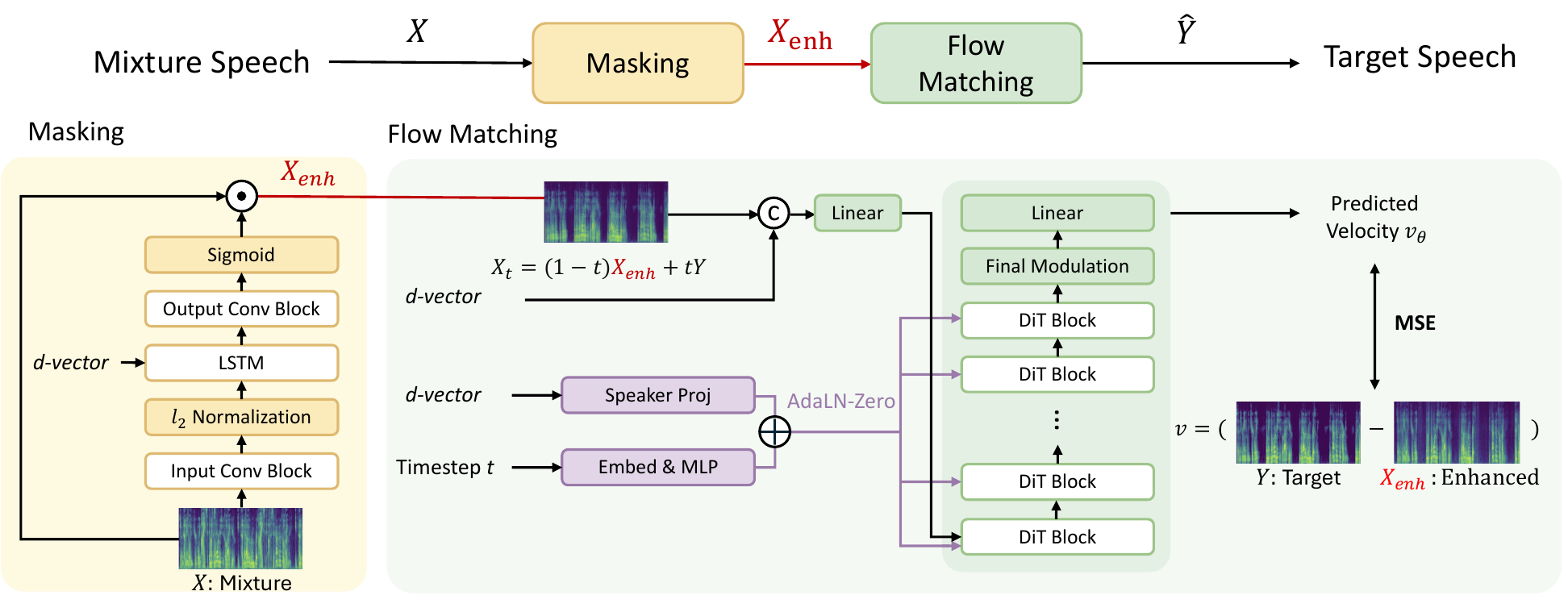}
  \vspace{-0.3cm}
    \caption{The proposed Mask2Flow-TSE architecture with masking and flow matching
   stages.}
  \label{fig:fm_overall}
\end{figure*}

Figure~\ref{fig:fm_overall} illustrates the overall architecture.

Given the input mixture mel-spectrogram $X$ and a target speaker embedding $d$, the masking stage estimates a soft mask:
\begin{equation}
  M = \text{Masking}(X, d).
\label{eq:Masking}
\end{equation}
The estimated mask $M$ is then element-wise multiplied with the mixture $X$ to produce the enhanced spectrogram $X_{\text{enh}}$:
\begin{equation}
    X_{\text{enh}} = X \odot M.
\end{equation}
The following flow matching stage refines this coarse output.
The key insight is that starting from the already-enhanced $X_{\text{enh}}$ rather than Gaussian noise significantly reduces the generation burden.
The final output is produced by:
\begin{equation}
  \hat{Y} = \text{FlowMatching}(X_{\text{enh}}, d),
  \label{eq:flowmatch}
\end{equation}
where $\hat{Y}$ approaches the clean target speech $Y$ with only a single inference step.
Note that our model operates on log-mel spectrograms compatible with Whisper ASR~\cite{radford2023robust}, enabling direct integration as a plug-and-play frontend. 
The following subsections describe each stage in detail.

\subsection{Stage 1: Masking}\label{ssec:stage1}

Masking module takes the mixture mel-spectrogram $X$ and speaker embedding $d$ as inputs.
The inputs are processed by convolutional layers followed by stacked bidirectional LSTM layers with residual connections.
To provide consistent target speaker guidance, the speaker embedding is concatenated at each LSTM layer.
A final linear layer with sigmoid activation predicts a soft mask $M \in [0, 1]$ as in Eq.~\eqref{eq:Masking}.
The masking module is trained to minimize the reconstruction error:
\begin{equation}
  \mathcal{L}_{\text{mask}} = || X_{\text{enh}} - Y ||^2.
  \label{eq:mask_loss}
\end{equation}

\subsection{Stage 2: Flow Matching}\label{ssec:stage2}

The second stage refines the coarse output from the Masking module using rectified flow matching~\cite{liu2022flow}.
Unlike conventional approaches that start from Gaussian noise $\epsilon \sim \mathcal{N}(0, I)$, we initialize the flow trajectory directly from the enhanced spectrogram $X_{\text{enh}}$.
Since the masking stage already removes the majority of interfering components, $X_{\text{enh}}$ lies close to the target $Y$, making the velocity field nearly constant along the trajectory.

In our formulation, the source $x_0$ in Eq.~\eqref{eq:interpolation} corresponds to $X_{\text{enh}}$ and the target $x_1$ to $Y$.
The flow trajectory is defined as a linear interpolation:
\begin{equation}
  X_t = (1-t)X_{\text{enh}} + tY,
  \label{eq:trajectory}
\end{equation}
where $t \in [0, 1]$ and $Y$ is the clean target spectrogram.

We adopt the Diffusion Transformer (DiT)~\cite{peebles2023scalable} and extend its AdaLN-Zero modulation for speaker-aware generation. Unlike the standard formulation that conditions solely on the timestep $t$, we explicitly modify the modulation to jointly condition on both $t$ and the speaker embedding $d$.
This efficiently injects speaker identity into every layer without cross-attention, while preserving the standard identity initialization.

The training objective for the flow matching model is to predict the velocity field $v_\theta$:
\begin{equation}
  \mathcal{L}_{\text{flow}} = \mathbb{E}_{t, X_{\text{enh}}, Y} || v_\theta(X_t, t, d) - (Y - X_{\text{enh}})||^2.
  \label{eq:flow_loss}
\end{equation}

At inference, starting from $X_{\text{enh}}$ at $t=0$, a single Euler step is sufficient to produce the final output:
\begin{equation}
  \hat{Y} = X_{\text{enh}} + v_\theta(X_{\text{enh}}, 0, d).
  \label{eq:single_step}
\end{equation}

\section{Experimental Setup}\label{sec:experiments}

\begin{figure}[t!]
    \centering
    \includegraphics[width=0.7\linewidth]{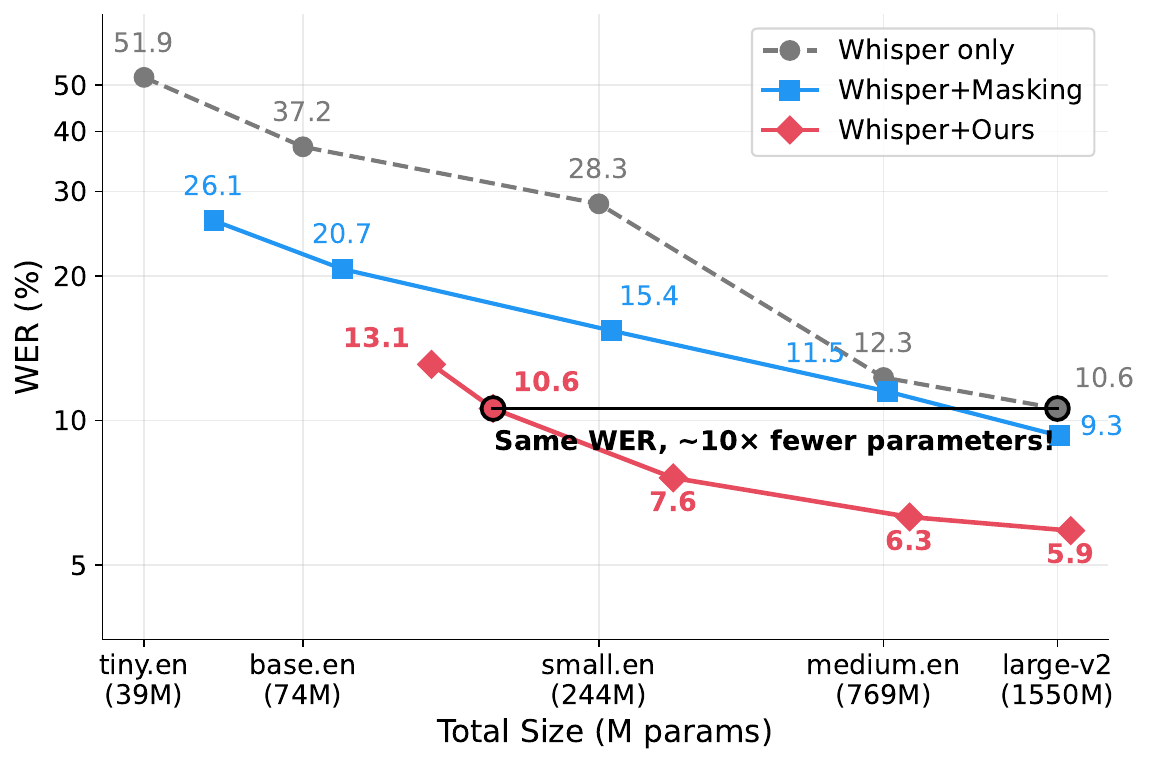}
    \caption{Comparison of WER and total system size under speech additive noise.
    Each point represents a (Whisper backbone, TSE method) pair.
    Our approach achieves the same WER as Whisper large-v2 alone with $\sim$10$\times$ fewer parameters.}
    \label{fig:wer_size}
    \vspace{0.2cm}
\end{figure}

\subsection{Datasets}\label{ssec:datasets}

We train our model on the Libri2Mix dataset~\cite{cosentino2020librimix}, a widely used TSE benchmark derived from LibriSpeech~\cite{panayotov2015librispeech} and WHAM!\ noise~\cite{wichern2019wham}, using both \texttt{mix\_clean} and \texttt{mix\_both} conditions from the training set.
To improve robustness to reverberant conditions and prevent overfitting to a single dataset, we additionally construct on-the-fly augmented mixtures from the full LibriSpeech training corpus, following the data creation protocol of VoiceFilter-Lite~\cite{wang2020voicefilter}.
In this setup, we consider three acoustically distinct conditions: \textbf{Clean}, where no augmentation is applied;
\textbf{Speech Additive}, where an interfering speaker of a different identity is mixed with the target speech;
and \textbf{Speech Reverb}, where the speech-additive mixture is convolved with room impulse responses (RIRs) from OpenSLR~\cite{ko2017study}.
For speech additive and reverb conditions, the signal-to-noise ratio (SNR) is uniformly sampled as $s \sim \mathcal{U}(1, 10)$\,dB.
All mixtures are fully overlapping, requiring the model to rely entirely on speaker characteristics for extraction.
Every audio is sampled at 16\,kHz, and the reference utterance is randomly selected from a different utterance of the same speaker.

Using the same on-the-fly protocol, we construct an evaluation set from the LibriSpeech \texttt{test-clean} and \texttt{test-other} subsets, covering clean, speech-additive, and reverberant conditions (Table~\ref{tab:custom_dataset}).
We additionally evaluate on the standard Libri2Mix benchmark~\cite{cosentino2020librimix} to enable direct comparison with prior work (Table~\ref{tab:librimix_summary}).

\begin{table*}[t!]
\centering
\setlength{\tabcolsep}{10pt}
\caption{WER~(\%) on LibriSpeech test-clean / test-other under three acoustic conditions. All models are evaluated with four Whisper ASR backbones. Params excludes the speaker encoder and Whisper backbone. Best WER is in \textbf{bold}.}
\resizebox{\linewidth}{!}{%
\scriptsize
\begin{tabular}{l|cc|cc|cc|c}
\toprule
\multirow{3}{*}{\rule{0pt}{4.5ex}\textbf{Model}}
& \multicolumn{2}{c|}{\multirow{2}{*}{\rule{0pt}{3.5ex}\textbf{Clean}}}
& \multicolumn{4}{c|}{\rule{0pt}{2.0ex}\textbf{Speech Noise}}
& \multirow{3}{*}{\rule{0pt}{4.5ex}\textbf{Params(M)}} \\
\cline{4-7}
& \multicolumn{2}{c|}{}
& \multicolumn{2}{c|}{\rule{0pt}{3.0ex}\textbf{Additive}}
& \multicolumn{2}{c|}{\rule{0pt}{3.0ex}\textbf{Reverb}}
& \\
\cline{2-7}
& t-clean & t-other & t-clean & t-other & t-clean & t-other & \\
\midrule
\multicolumn{8}{c}{\textbf{\texttt{Whisper tiny.en}}} \\
\midrule
No preprocessing      & 5.7 & 14.5 & 51.9 & 73.8 & 65.8 & 89.4 & N/A \\
\midrule
ConVoiFilter~\cite{nguyen2024convoifilter}           & 13.0 & 19.8 & 25.5 & 39.0 & 48.8 & 61.1 & 49.9 \\
TSELM~\cite{tang2025tselm}                 & 10.8 & 22.7 & 13.8 & 27.5 & 44.8 & 58.2 & 195.4 \\
Metis-TSE~\cite{wang2025metis}              & 9.7 & 23.2 & 14.3 & 29.0 & 49.3 & 58.9 & 1425 \\
\midrule
\textbf{Mask2Flow-TSE} & \textbf{5.7} & \textbf{14.1} & \textbf{13.1} & \textbf{27.4} & \textbf{33.3} & \textbf{49.6} & 85.3 \\
\midrule
\multicolumn{8}{c}{\textbf{\texttt{Whisper base.en}}} \\
\midrule
No preprocessing      & 4.3 & 10.4 & 37.2 & 48.4 & 49.4 & 66.8 & N/A \\
\midrule
ConVoiFilter~\cite{nguyen2024convoifilter}           & 11.2 & 16.0 & 21.3 & 31.2 & 40.8 & 52.1 & 49.9 \\
TSELM~\cite{tang2025tselm}                 & 9.1 & 19.0 & 12.1 & 24.1 & 41.4 & 55.2 & 195.4 \\
Metis-TSE~\cite{wang2025metis}              & 7.7 & 18.4 & 12.2 & 24.5 & 47.3 & 57.6 & 1425.0 \\
\midrule
\textbf{Mask2Flow-TSE} & \textbf{4.3} & \textbf{10.8} & \textbf{10.6} & \textbf{21.0} & \textbf{28.1} & \textbf{42.2} & 85.3 \\
\midrule
\multicolumn{8}{c}{\textbf{\texttt{Whisper small.en}}} \\
\midrule
No preprocessing      & 3.1 & 7.4 & 28.3 & 38.6 & 40.0 & 55.9 & N/A \\
\midrule
ConVoiFilter~\cite{nguyen2024convoifilter}           & 10.3 & 12.8 & 17.2 & 24.1 & 33.7 & 42.0 & 49.9 \\
TSELM~\cite{tang2025tselm}                 & 7.5 & 14.5 & 9.4 & 19.5 & 38.9 & 51.6 & 195.4 \\
Metis-TSE~\cite{wang2025metis}              & 5.7 & 14.1 & 10.3 & 20.4 & 44.9 & 52.9 & 1425.0 \\
\midrule
\textbf{Mask2Flow-TSE} & \textbf{3.1} & \textbf{7.4} & \textbf{7.6} & \textbf{16.1} & \textbf{23.4} & \textbf{34.7} & 85.3 \\
\midrule
\multicolumn{8}{c}{\textbf{\texttt{Whisper medium.en}}} \\
\midrule
No preprocessing      & 3.0 & 5.8 & 12.3 & 22.2 & 21.0 & 33.4 & N/A \\
\midrule
ConVoiFilter~\cite{nguyen2024convoifilter}               & 10.2 & 11.3 & 14.9 & 20.6 & 29.8 & 37.1 & 49.9 \\
TSELM~\cite{tang2025tselm}             & 6.4 & 12.5 & 8.7 & 17.7 & 36.5 & 49.5 & 195.4 \\
Metis-TSE~\cite{wang2025metis}              & 5.1 & 12.2 & 9.4 & 18.0 & 44.1 & 50.6 & 1425.0 \\

\midrule
\textbf{Mask2Flow-TSE} & \textbf{2.6} & \textbf{5.8} & \textbf{6.3} & \textbf{13.1} & \textbf{19.8} & \textbf{30.1} & 85.3 \\
\bottomrule
\end{tabular}
\vspace{0.2cm}
}
\label{tab:custom_dataset}
\end{table*}

\subsection{Baselines}\label{ssec:baselines}
We compare Mask2Flow-TSE against both discriminative and generative TSE methods.
For discriminative approaches, we include ConVoiFilter~\cite{nguyen2024convoifilter}, a convolutional architecture with X-vector speaker embeddings.
For generative methods, we consider TSELM~\cite{tang2025tselm}, which employs a non-autoregressive transformer to predict WavLM-derived discrete tokens; and Metis-TSE~\cite{wang2025metis}, a large-scale DiT-based model operating on discrete audio codec tokens.

\subsection{Evaluation Metrics}\label{ssec:metrics}
Since our primary goal is to improve downstream ASR performance, we adopt word error rate (WER) as the main evaluation metric.
We measure WER using Whisper~\cite{radford2023robust} for Table~\ref{tab:custom_dataset} and Table~\ref{tab:librimix_summary}.

To evaluate speaker identity preservation, we compute Speaker Similarity (SpkSim) as the cosine similarity between speaker embeddings extracted from waveforms reconstructed with the original phase using a pretrained WeSpeaker model~\cite{wang2023wespeaker}.
We also report the Real-Time Factor (RTF), defined as the ratio of processing time to audio duration, measuring only the TSE model inference time on a single NVIDIA V100 16GB GPU with a batch size of 1.  
\section{Results}\label{sec:results}

\begin{table*}[t!]
\centering
\caption{TSE results on the Libri2Mix test set using the max configuration.
WER is measured with Whisper \texttt{large-v2}.
Best results are in \textbf{bold} and second-best are \underline{underlined}.}
\resizebox{\linewidth}{!}{%
\setlength{\tabcolsep}{13pt}%

\begin{tabular}{l|cc|cc|cc|c}
\toprule
\multirow{2}{*}{\textbf{Model}} 
& \multicolumn{2}{c|}{\textbf{Libri2Mix Noisy}}
& \multicolumn{2}{c|}{\textbf{Libri2Mix Clean}}
& \multicolumn{2}{c|}{\textbf{Single Speaker}} 
& \multirow{2}{*}{\textbf{RTF}$\downarrow$} \\
& WER$\downarrow$ & SIM$\uparrow$ & WER$\downarrow$ & SIM$\uparrow$ & WER$\downarrow$ & SIM$\uparrow$ & \\
\midrule
No processing & 67.8 & 0.45 & 63.2 & 0.54  & 2.9 & 1.0 & - \\
\midrule
ConVoiFilter~\cite{nguyen2024convoifilter}  & \textbf{37.5} & \underline{0.56} & 18.0 & \underline{0.78} & 10.5 & \textbf{0.89} & \textbf{0.004} \\
TSELM~\cite{tang2025tselm}                  & 42.1 & 0.23 & 24.3 & 0.25 & 8.9 & 0.27 & 0.172 \\
Metis-TSE~\cite{wang2025metis}               & 43.9 & \textbf{0.58} & \underline{16.4} & 0.61 & \underline{5.1} & 0.62 & 1.427 \\
\midrule
\textbf{Mask2Flow-TSE}              & \underline{39.6} & \underline{0.56} & \textbf{13.2} & \textbf{0.80} & \textbf{2.9} & \underline{0.83} & \underline{0.007} \\
\bottomrule
\end{tabular}}
\vspace{0.2cm}
\label{tab:librimix_summary}
\end{table*}

\subsection{WER Evaluation on LibriSpeech}\label{ssec:main_results}

Table~\ref{tab:custom_dataset} compares WER across four models under three acoustic conditions.
In the clean condition, Mask2Flow-TSE preserves the original speech quality with minimal WER degradation across all Whisper model variants.
In contrast, all baselines noticeably increase WER on clean speech, suggesting that they implicitly assume a mixture scenario and apply unnecessary processing even when no interference is present.

Under speech noise conditions (both additive and reverberant), Mask2Flow-TSE achieves the lowest WER across all Whisper models and both test subsets, outperforming TSELM and Metis-TSE despite using significantly fewer parameters.
This advantage is maintained consistently as the ASR model scales up, whereas other models show diminishing gains with stronger recognizers.

Figure~\ref{fig:wer_size} visualizes the WER--model size trade-off under speech additive noise.
Notably, Mask2Flow-TSE with Whisper \texttt{base.en} (159.3M total) achieves the same WER (10.6\%) as Whisper \texttt{large-v2} alone (1550M) with approximately 10$\times$ fewer parameters.

\subsection{TSE Evaluation on Libri2Mix}\label{ssec:librimix}

Table~\ref{tab:librimix_summary} shows WER, speaker similarity, and RTF on the Libri2Mix test set.
Mask2Flow-TSE achieves the best WER in the clean condition and competitive performance under noisy conditions, while preserving the original WER on single-speaker utterances, where all baselines suffer from degradation.

Furthermore, as a single-step generative model, Mask2Flow-TSE achieves an RTF comparable to the masking-based model (ConVoiFilter) and orders of magnitude faster than other generative baselines.
Overall, Mask2Flow-TSE achieves both the speed of masking-based models and the quality of generative approaches.

\section{Analysis}
\label{sec:analysis}

\subsection{Delete--Insert Analysis}
In Section~\ref{ssec:motivation}, we show that flow-only TSE models spend most of their inference budget on deletion. Here, we verify that our two-stage pipeline redistributes this budget as intended.

Table~\ref{tab:di_analysis} reports the D/I proportion at each stage. While the masking stage performs pure deletion, the flow matching stage operating on the masked output is dominated by insertion (I\,$=$\,83.3\% on clean, 61.8\% on noisy), restoring spectral details that masking over-suppresses. The lower insertion ratio under the noisy condition reflects additional removal of residual noise.

Notably, the overall D/I proportion of Mask2Flow-TSE closely matches the ground-truth target in both conditions, demonstrating that the model captures the correct deletion-insertion proportion.
By offloading deletion to the masking stage, the flow matching module can focus on insertion and achieve accurate generation in a single step.

\begin{table}[t]
\centering
\begin{minipage}[t]{0.48\linewidth}
\vspace{0pt}
\centering
\caption{D/I energy proportion at each stage on Libri2Mix (Eq.~\eqref{eq:di_energy}).}
\vspace{0.2cm}
\label{tab:di_analysis}
   \resizebox{1.0\linewidth}{!}{
   \renewcommand{\arraystretch}{1.1}
    \begin{tabular}{rcl | cc | cc}
    \toprule
    \multirow{2}{*}{Input} & \multirow{2}{*}{$\rightarrow$} & \multirow{2}{*}{Output} & \multicolumn{2}{c|}{Noisy} & \multicolumn{2}{c}{Clean} \\
                           &                                &                         & D(\%) & I(\%) & D(\%) & I(\%) \\
    \midrule
    Mixture & $\rightarrow$ & Masked      & 100.0 & 0.0           & 100.0 & 0.0           \\
    Masked  & $\rightarrow$ & Mask2Flow   & 38.2  & \textbf{61.8} & 16.7  & \textbf{83.3} \\
    \midrule
    Mixture & $\rightarrow$ & Mask2Flow   & 80.3  & 19.7          & 77.2  & 22.8          \\
    Mixture & $\rightarrow$ & Target (GT) & 75.1  & 24.9          & 72.5  & 27.5          \\
    \bottomrule
  \end{tabular}}
\end{minipage}
\hfill
\begin{minipage}[t]{0.48\linewidth}
\vspace{0pt}
\centering
\caption{Effect of flow matching prior (LibriSpeech test-clean, speech noise).}
\vspace{0.2cm}
\label{tab:ablation_prior}
   \resizebox{1.0\linewidth}{!}{
   \renewcommand{\arraystretch}{1.15}
    \begin{tabular}{lcccc}
    \toprule Method    & Prior ($\mathbf{x}_{0}$)              & Condition & WER$\downarrow$ & Steps      \\
    \midrule Flow-only & $\mathcal{N}(\mathbf{0}, \mathbf{I})$ & Mixture   & 13.7                & 16         \\
    Flow-only          & Mixture                               & N/A       & 12.2                & 8          \\
    Mask2Flow      & Masked                                & N/A       & \textbf{7.6}        & \textbf{1} \\
    \bottomrule
  \end{tabular}}
\end{minipage}

\end{table}

\subsection{Ablation Study}
\label{ssec:ablation}

A key design choice in Mask2Flow-TSE is using the masked mel spectrogram as the flow
matching prior, rather than sampling from Gaussian noise
$\mathbf{z}\sim \mathcal{N}(\mathbf{0}, \mathbf{I})$ as in conventional flow-based TSE methods~\cite{navon2025flowtse}. Using the same setup as Table~\ref{tab:custom_dataset}
(LibriSpeech test-clean, speech additive noise, Whisper \texttt{small.en}), Table~\ref{tab:ablation_prior}
compares three prior choices while keeping the architecture identical.

The results show that a prior closer to the target clean speech significantly reduces
inference steps and WER. While a Gaussian prior requires the longest trajectory (16
steps) to learn source separation from scratch, using a mixture prior halves the
steps by providing a speech-like starting point.
Mask2Flow-TSE further achieves the best WER in a single step by leveraging the masked mel spectrogram.

\begin{wraptable}[11]
  {r}{0.45\textwidth}
  \vspace{-12pt}
  \centering
  \setlength{\tabcolsep}{4pt}
  \caption{Ablation study of model architecture on WER (\%).}
  \label{tab:ablation_mask2flow}
  \resizebox{\linewidth}{!}{%
  \scriptsize
  \begin{tabular}{l|cc|cc}
    \toprule \multirow{3}{*}{\rule{0pt}{3.0ex}\textbf{Model}} & \multicolumn{4}{c}{\rule{0pt}{1ex}\textbf{Speech Noise}} \\
    \cline{2-5}                                               & \multicolumn{2}{c|}{\rule{0pt}{3.0ex}\textbf{Additive}} & \multicolumn{2}{c}{\rule{0pt}{3.0ex}\textbf{Reverb}} \\
    \cline{2-5}                                               & t-clean                                                 & t-other                                             & t-clean       & t-other       \\
    \midrule No preprocessing                                 & 28.3                                                    & 38.6                                                & 40.0          & 55.9          \\
    \midrule Masking                                          & 15.4                                                    & 25.3                                                & 33.0          & 46.2          \\
    Flow-only                                                 & 12.2                                                    & 19.6                                                & 25.7          & 37.2          \\
    \textbf{Mask2Flow}                                    & \textbf{7.6}                                            & \textbf{16.1}                                       & \textbf{23.4} & \textbf{34.7} \\
    \bottomrule
  \end{tabular}
  \vspace{0.2cm}
  }
\end{wraptable}

Table~\ref{tab:ablation_mask2flow}, which additionally includes reverberation, demonstrates the effectiveness of the proposed architecture. While flow-only models outperform simple masking, their performance is constrained by
the difficulty of modeling the full transformation from the mixture to the target speech
through iterative flow steps.
By introducing a lightweight 12.7M masking module, Mask2Flow-TSE
alleviates this burden on the flow-matching stage and achieves superior
performance with a single Euler step compared with a 72.6M flow-only model
using eight steps.

\section{Conclusion}\label{sec:conclusion}
In this work, we proposed Mask2Flow-TSE, a principled two-stage framework that combines discriminative masking with generative flow matching for target speaker extraction. The design is motivated by our newly introduced deletion/insertion (D/I) proportion analysis, which reveals that deletion dominates the early steps of flow inference.
Based on this observation, Mask2Flow-TSE offloads deletion to a lightweight masking network,
allowing the flow-matching stage to focus on generative refinement and achieve accurate reconstruction with only a single Euler step.
Experimental results demonstrate that explicitly decomposing TSE into discriminative deletion and generative refinement enables robust recognition performance with substantially lower inference cost and model complexity.
Beyond target speaker extraction, the
Mask2Flow paradigm suggests a general strategy for speech processing tasks that require
both the removal of undesired components and the restoration of desired speech content,
such as speech enhancement, dereverberation, and bandwidth extension.

% \begin{ack}  % only for camera-ready
% Acknowledgement 
% \end{ack}

{
    \small
    \bibliographystyle{plain}
    \bibliography{references}
}

%%%%%%%%%%%%%%%%%%%%%%%%%%%%%%%%%%%%%%%%%%%%%%%%%%%%%%%%%%%%

\newpage
\appendix
\section{Appendix}

\subsection{Model Configurations}\label{ssec:model_config}

\subsubsection{Speaker Encoder}
We employ WavLM-base-plus-sv~\cite{chen2022wavlm} pretrained on speaker verification as the speaker encoder.
It takes a raw waveform as input and extracts a 512-dimensional d-vector from the reference utterance.
The speaker encoder remains frozen during training, and the extracted d-vector is shared across both the masking and flow matching stages.

\subsubsection{Masking}
The input 80-dim log mel-spectrogram is $\ell_2$-normalized and processed by a stack of four Conv2d layers with batch normalization and LeakyReLU activation, producing 8-channel feature maps.
These features are flattened and concatenated with the 512-dimensional speaker embedding, then fed into a 2-layer bidirectional LSTM with 416 hidden units.
The speaker embedding is concatenated at each LSTM layer to provide consistent target speaker guidance.
LayerNorm and residual connections are applied after each layer.
The output is passed through a linear layer with sigmoid activation to produce the soft mask.

\subsubsection{Flow Matching}
The flow matching module operates on 80-dim log mel-spectrograms and employs a DiT (Diffusion Transformer) backbone with 9 blocks, 768-dimensional hidden states, and 8 attention heads.
We use rotary position embeddings (RoPE)~\cite{su2024roformer} for sequence modeling.
The speaker embedding is incorporated in two ways: it is concatenated with the input features and the enhanced spectrogram, and it is also injected into the AdaLN-Zero conditioning jointly with the timestep embedding.
This dual conditioning enables speaker-aware generation at every transformer layer without requiring a separate cross-attention mechanism, reducing computational cost while maintaining strong speaker fidelity.
At inference, we use the Euler ODE solver with a single integration step, which is sufficient due to the straight-path property of rectified flow matching.

\subsubsection{Training}
We adopt a sequential training scheme.
First, the masking module is trained with $\mathcal{L}_{\text{mask}}$ (Eq.~\eqref{eq:mask_loss}) until convergence.
Then, the masking module is frozen, and the flow matching module is trained with $\mathcal{L}_{\text{flow}}$ (Eq.~\eqref{eq:flow_loss}).
All audio is segmented into 10-second chunks at 16\,kHz during training, with a batch size of 10 and 2-step gradient accumulation.
We use the AdamW~\cite{loshchilov2017decoupled} optimizer with a learning rate of $2 \times 10^{-4}$ and 10k warmup iterations.
Exponential moving average (EMA) with decay 0.9999 is applied for stable inference.
The total model size is approximately 85M parameters: 12.7M for Masking, 72.6M for Flow Matching.

\subsubsection{Compute Resources}
All experiments are conducted on a single NVIDIA V100 GPU with 16GB memory.
Inference latency (Real-Time Factor, Section~\ref{ssec:metrics}) is measured on a single V100 with batch size 1.

\section{More Analysis}

\subsection{Speech Enhancement vs.\ Target Speaker Extraction}\label{ssec:se_vs_tse}

Speech enhancement (SE) suppresses background noise and reverberation under the assumption of a single dominant speaker, without using speaker-specific information.
In contrast, target speaker extraction (TSE) addresses the cocktail-party scenario where multiple speakers may be present, and explicitly extracts the voice of a designated target speaker from the mixture.
This requires auxiliary speaker information, which is conventionally provided as a short reference (enrollment) utterance from the target speaker.
The reference is encoded into a speaker embedding via a pretrained speaker verification model and serves as a conditioning input throughout the extraction process.
We follow this standard TSE setup in this work.

\subsection{Why Masking Requires a Mixture Prior}
\begin{figure}[t!]
    \centering
    \includegraphics[width=0.6\linewidth]{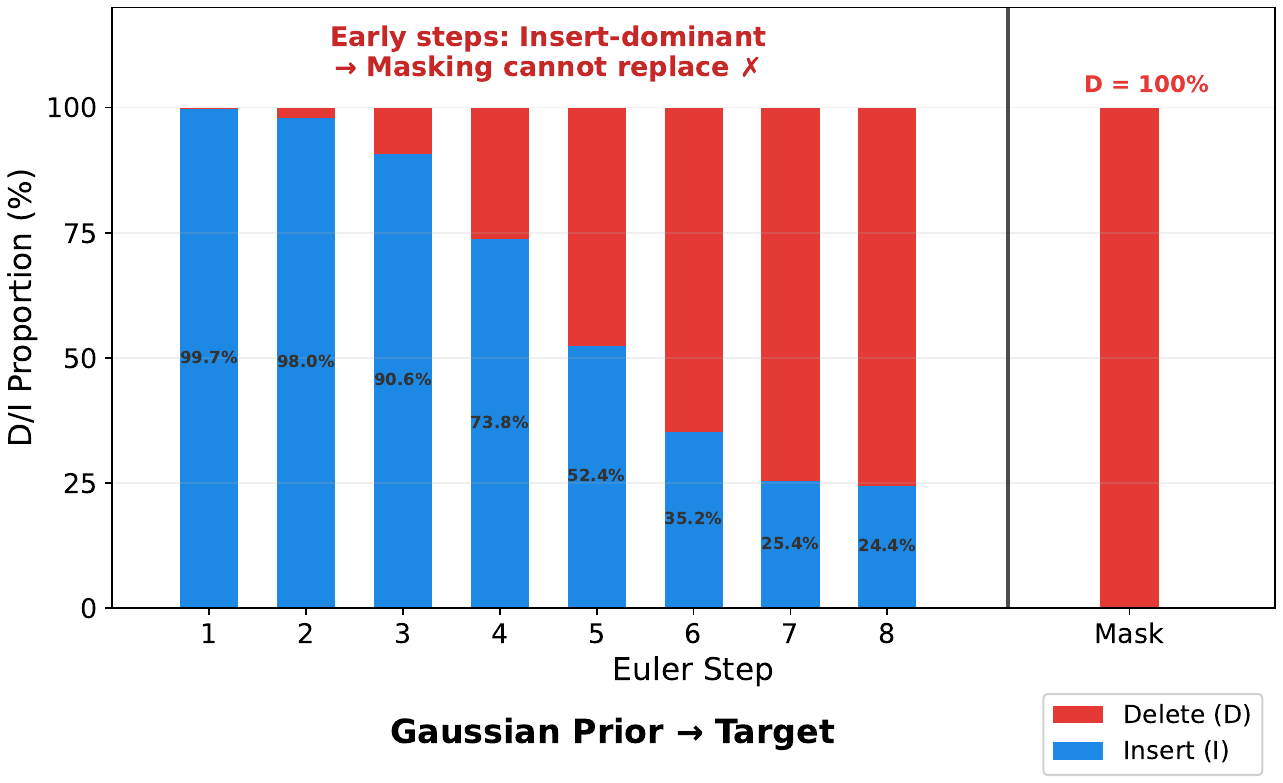}
    \caption{D/I energy proportion per Euler step for a Gaussian-prior flow model on Libri2Mix clean.
    Early steps are insert-dominant, opposite to masking (D\,$=$\,100\%).}
    \label{fig:gaussian_vs_mixture_di}
\end{figure}

Figure~\ref{fig:gaussian_vs_mixture_di} further illustrates
the prior choice from the D/I perspective.
When starting from Gaussian noise, the first Euler step
is almost entirely insertion (I\,$=$\,99.7\%),
and the transformation remains insert-dominant
through step~5 (I\,$=$\,52.4\%).
Since masking is a pure deletion operation (D\,$=$\,100\%),
it operates in the opposite direction to these early steps
and therefore cannot serve as a substitute for them.
In contrast, Figure~\ref{fig:motivation}(b) shows that
when starting from the mixture,
every step is deletion-dominant from the outset
(D\,$=$\,94.1\% at step~1),
closely aligned with masking.
This alignment is precisely what enables our two-stage design:
the masking stage can effectively replace
the deletion-heavy early steps of the flow,
allowing the flow matching module to focus exclusively
on the remaining insertion.

\begin{figure}[t!]
     \centering
     \includegraphics[width=0.75\linewidth]{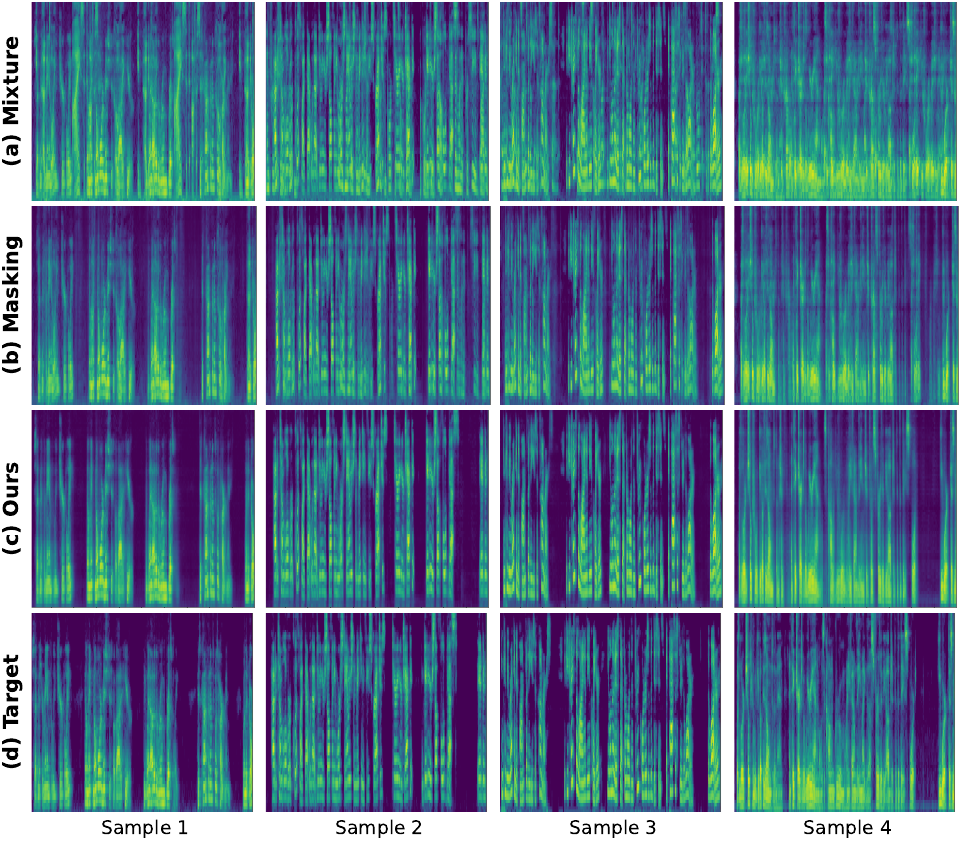}
     \caption{Spectrogram comparison on Libri2Mix test samples. From top to bottom: (a) input mixture, (b) masking stage output, (c) Mask2Flow-TSE output, and (d) clean target.}
     \label{fig:fm_spec}
\end{figure}

\subsection{Spectrogram Analysis}\label{ssec:spectrogram}

Figure~\ref{fig:fm_spec} provides a qualitative comparison of spectrograms at each stage of our pipeline on four Libri2Mix test samples.
The mixture~(a) shows overlapping spectral energy from both speakers across the entire frequency range, making the target speech difficult to distinguish.

First, the masking stage~(b) effectively suppresses the interfering speaker but introduces visible over-suppression of the target speech.
This is most apparent in the low-frequency harmonic regions (below $\sim$2\,kHz), where the periodic vertical striations corresponding to voiced speech are noticeably faded compared to the clean target~(d).
As a result, the masked spectrograms appear overly smooth and lack fine spectral detail, particularly in Samples~1--3.

Second, the flow matching stage~(c) restores these over-suppressed harmonic structures, recovering the sharpness and periodicity of the target speaker's voiced segments.
In Samples~1 and~3, the harmonic patterns in~(c) closely match the clean target~(d), whereas the masking output~(b) shows substantial energy loss in these regions.
Furthermore, in Sample~4, where the target speaker is silent during the latter portion of the utterance, the flow matching stage preserves silence without hallucinating spurious spectral content, suggesting that the model does not blindly insert energy but selectively restores only the missing target components.

\section{Limitations}\label{ssec:limitations}

While Mask2Flow-TSE reduces parameters and inference latency, it still requires a pre-trained speaker encoder.
For strict on-device deployment, minimizing the size and computational overhead of this encoder remains a key area for future work.
Furthermore, the current model prioritizes efficiency over capacity.
Scaling it into a unified, multi-task speech preprocessing model, following recent unified frontends such as Metis~\cite{wang2025metis} and UniFlow~\cite{wang2025uniflow}, remains a promising direction for future work.

%%%%%%%%%%%%%%%%%%%%%%%%%%%%%%%%%%%%%%%%%%%%%%%%%%%%%%%%%%%%

\newpage

\end{document}